\documentclass[]{cjaa}           

\usepackage{graphicx}                   
\input{epsf.sty}                        
\input{psfig.sty}                       

\begin{document}

   \title{Microquasars and gamma-ray sources
}

   \volnopage{Vol.0 (200x) No.0, 000--000}      
   \setcounter{page}{1}          

   \author{Gustavo E. Romero
      \inst{1,2,}\mailto{}
        }
   \offprints{G.E. Romero}                   

   \institute{Instituto Argentino de Radioastronom\'{\i}a, C.C. 5, 1894 Villa Elisa, Argentina\\
             \email{romero@irma.iar.unlp.edu.ar}
        \and
             Facultad de Ciencias Astron\'omicas y Geof\'{\i}sicas, UNLP, Paseo del Bosque, 1900 La Plata, Argentina\\
             \email{romero@venus.fisica.unlp.edu.ar}
          }

   \date{Received~~2004 month day; accepted~~2004~~month day}

   \abstract{Microquasars are X-ray binary systems with non-thermal radio emission originated in jet-like features. They are attractive sites for gamma-ray production, since relativistic particles in the jet should traverse locally strong both photon and matter fields. In this review we discuss whether some of the unidentified gamma-ray sources detected by the EGRET instrument of the Compton Gamma-Ray Observatory might be associated with microquasars. Relevant models for gamma-ray emission in such binaries are described and prospects for the detection of microquasars with instruments of new generation are briefly evaluated.
   }

   \authorrunning{G.E. Romero }            
   \titlerunning{Microquasars and $\gamma$-ray sources}  

   \maketitle

%
%
\section{Introduction}           
\label{sect:intro}


Microquasars are accreting compact objects in X-ray binaries that produce jets of relativistic particles (Mirabel \& Rodr\'{\i}guez 1999). The behavior of these systems resembles, to some extent, that presented by extragalactic quasars, hence the origin of the name. There are, nonetheless, important differences. One is the presence of a companion star which provides the mass that accretes onto the compact object. According to the nature of the mass donor, microquasars can be classified into two flavors: high-mass microquasars, where the companion is a young massive star and the mass transfer occurs through the stellar wind, and low-mass microquasars, where the companion is an older star which transfers mass by Roche lobe overflow. The compact object can be either a black hole or a weakly magnetized neutron star. The number of known microquasars is $\sim 15$ at present, but the actual population is surely much higher, perhaps up to represent $\sim 70$ \% of all X-ray binaries (Fender \& Maccarone 2004).  

Black hole microquasars are found in two different spectral states at X-ray energies. In the `hard' state, they present a power-law X-ray emission extending up to $\sim 100$ keV. This emission is thought to be originated in a hot corona around the black hole that cools by Comptonization of soft X-ray photons from the colder disk (e.g. Poutanen 1998,  McClintock \& Remillard 2004). During this state a steady, self-absorbed jet is observed at radio wavelengths. In the `soft' state the X-ray emission is dominated by the disk component and the radio jet is absent. The transition between both states occurs at very high luminosities and is associated with the formation of discrete jets (Fender \& Maccarone 2004). Neutron star binaries show a similar behavior, also with relativistic ejections (Fender et al. 2004), but lower radio luminosities.    

Some microquasars, like LS 5039 (Paredes et al. 2002) and LSI +61 303 (Massi et al. 2004), seem to elude the above spectral behavior and show persistent jets and low X-ray luminosities all the time. The jets are short and of moderate power. Both mentioned cases are suspected to harbor a neutron star as the compact object. 

Microquasars have all the elements to produce gamma-rays: they have outflows of relativistic particles with mild bulk motions, strong photon fields that can act as a target for energetic electrons and positrons, and also (in the case of high-mass microquasars) dense matter fields in the form of stellar winds that could interact with relativistic hadrons to produce $\pi^0$ gamma-rays and neutrinos from charged pion decays. Other astrophysical systems with similar ingredients, like blazars, have been detected at very high energies. It is natural, then, to consider microquasars as potential counterparts for some of the unidentified gamma-ray sources detected by the EGRET instrument of the Compton Gamma-Ray Observatory. 

In this article we will review the main phenomenology of the EGRET sources and we will discuss the physical mechanisms that could generate gamma-ray emission in microquasars. We will then ponder whether some of the unidentified sources might correspond to microquasars yet undetected at other wavelengths.     

\section{Phenomenology of gamma-ray sources}
\label{sect:Phe}

The Compton Gamma-Ray Observatory was launched in early 1991 and deorbited in June, 2000. Between April 1991 and September 1995, the Energetic Gamma Ray Experiment Telescope (EGRET) detected 416 gamma-ray excesses with a significance of more than 3 $\sigma$ above the diffuse background emission. After a careful analysis with a maximum-likelihood procedure, only 271 of these sources were included in the final official list published as the Third EGRET (3EG) Catalog (Hartman et al. 1999). The included detections have a significance of more than 5 $\sigma$ at less than 10 degrees from the galactic plane, where the background radiation is stronger, and more than 4 $\sigma$ otherwise. Single power-law spectra can be fitted to most sources, and the corresponding spectral photon index usually ranges between 1.5 and 3.0. More than 150 of these sources remain yet unidentified.

A simple analysis of the distribution of the unidentified EGRET sources with galactic coordinates reveals that there is a clear concentration of sources on the galactic plane plus a concentration in the general direction of the galactic center (see Figures 1 and 2). This indicates a significant contribution from galactic sources. If many or most of the low-latitude gamma-ray sources are produced by galactic objects, one of the first things we could ask is whether they are correlated with the spiral arms of the Galaxy, i.e. the places where stars are formed. A correlation analysis between 3EG sources and bright and giant HII regions (the usual tracer for the galactic spiral structure) shows that there is a strong correlation at $\sim 7\sigma$-level (Romero 2001). This means that there is a significant number of Population I objects in the parent population of the low-latitude gamma-ray sources. 

\begin{figure}[h]
 \begin{minipage}[t]{0.5\linewidth}
\centering
  \includegraphics[width=65mm,height=65mm]{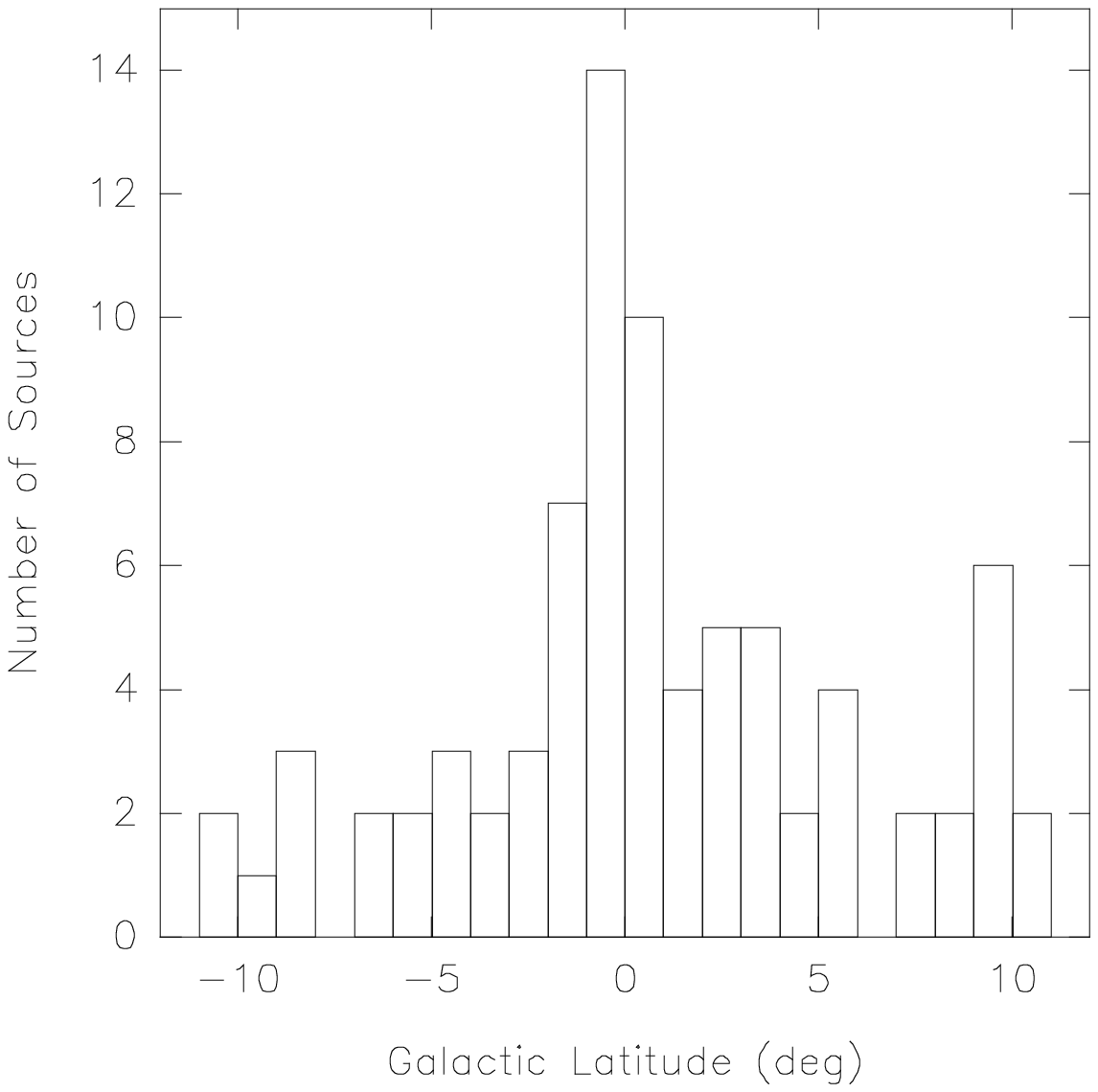}
 \vspace{-5mm}
\caption{{\small Distribution of EGRET sources with galactic latitude. From Romero et al. (1999).} }
 \end{minipage}%
 \begin{minipage}[t]{0.5\textwidth}
 \centering
 \includegraphics[width=60mm,height=65mm]{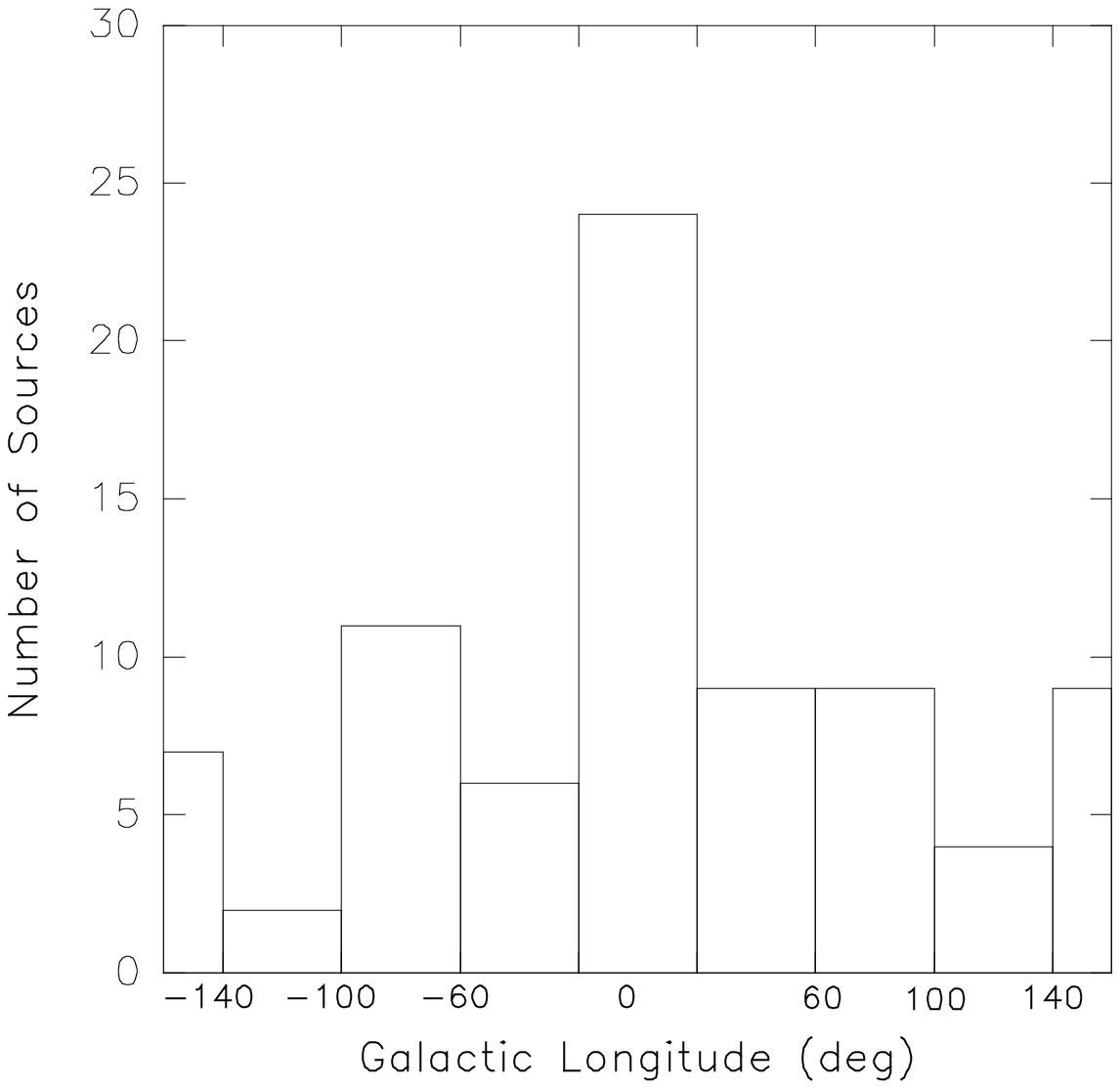}
 \vspace{-5mm}
 \caption{{\small Distribution of EGRET sources with galactic longitude. From Romero et al. (1999).}}
 \end{minipage}%
 \label{Fig:f12}
\end{figure}

Because of the large error boxes in the location of the EGRET sources (typically $\sim 1$ degree in diameter), correlation studies have a limited potential to find secure counterparts. This technique has been successful mainly to identify bright blazars at high latitudes, where the confusion is significantly lower than in the galactic plane. A complementary approach is provided by population studies using the known characteristics of the gamma-ray sources, like their spectra, variability and flux density. The so-called $\log N - \log S$ studies can be particularly useful on this respect (e.g. Reimer \& Thompson 2001). A $\log N - \log S$ plot displays the number of sources ($N$) with a gamma-ray flux ($S$) greater than a given value. By means of such an analysis of the steady sources in the 3EG catalog, Gehrels et al. (2000) have demonstrated the existence of at least two well-defined populations: one, a group of bright sources at low latitudes with an average spectral index of $2.18\pm0.04$, and the other, a group of weaker mid-latitude sources with softer spectra (an average index of $2.49\pm0.04$). The transition between both groups occurs at $|b|\sim 5$ degrees, i.e. far from where the source significance changes from $4\sigma$ to $5\sigma$. The low-latitude sources are probably young objects located in the inner spiral arms (see also Bhattacharya et al. 2003), whereas the mid-latitude sources might be nearby (100-400 pc) sources associated to the Gould Belt, a local star forming region (Grenier 1995, 2000).

More recent and complete population studies including variability and source distribution models that take into account the non-uniform detection sensitivity across the sky (Grenier 2001, 2004) suggest that there exist at least 4 different populations of gamma-ray sources: 1) Bright and relatively hard (photon index $\Gamma\sim 2.5$) sources near the plane ($|b|<5^{\circ}$), some of them variable. 2) Weaker and non-variable sources with $\Gamma\sim2.25$ that are spatially correlated with the Gould Belt. 3) A group of luminous sources with very soft spectra ($\Gamma\sim2.5$) and high variability forming a kind of halo around the galactic center with a scale height of $\sim2$ kpc. 4) An isotropic population of extragalactic origin and a variety of spectra and variability behaviors. There are no more than 35 sources in this group.

Regarding the galactic sources, the first group should be formed by young sources (a few million years at most) with isotropic luminosities in the range $10^{34-36}$ erg/s. These sources contain a subgroup of clearly variable sources (Torres et al. 2001, Nolan et al. 2003, Bosch-Ramon et al. 2004). The group 2 should be composed also by young sources but with luminosities in the range $10^{32-33}$ erg/s. The sources in the third group might be formed in and ejected from globular clusters (however, there is no significant correlation with individual clusters) or from the galactic plane. These sources should be old (age measured in Gyrs) and very luminous, in the range $10^{35-37}$ erg/s. Significant variability between different EGRET viewing periods is observed in many of these sources (Nolan et al. 2003).

\section{Microquasars as high-energy sources}
\label{sect:micro}

The discovery of the microquasar LS 5039 by Paredes et al. (2000) triggered a renewed interest in microqusars as potential gamma-ray sources. Until then, some theoretical work has been done exploring potential Synchrotron self-Compton (SSC) high-energy emission associated with non-thermal flares in SS433 (Band \& Grindlay 1986), GRS 1915+105 (Aharonian \& Atoyan 1998, Atoyan \& Aharonian 1999) and GRO 1655+40 (Levinson \& Blandford 1996). Paredes and collaborators proposed the association of LS 5039 with the unidentified EGRET source 3EG J1824-1514 and suggested that the gamma-ray emission might be the result of inverse Compton upscattering of UV seed photons coming from the stellar companion. Subsequent work by other authors (see below) expanded this proposal by including other photon fields and more sophisticated jet models. 

Recently, the discovery of radio jets in LSI +61 303 confirmed the microquasar nature of this binary (Massi et al. 2001, 2004). The correlation of the gamma-ray variability of the coincident gamma-ray source 3EG J0241+6103 with the orbital period of the binary seems to support the suggestion that LSI + 61 303 is another gamma-ray emitting microquasar (Massi 2004). A case for a third object has been recently presented by Combi et al. (2004), on the basis of the suggested association between AX J1639.0-4642 (actually a microquasar candidate at present) and 3EG J1639-4702.

Since microquasars seem to be capable of producing gamma-rays, at least in some cases with persistent jets and high-mass companions, and they are compact objects which can display significant variability at all wavelengths on short timescales, it is natural to suggest that the group of variable unidentified EGRET sources at low galactic latitudes could be formed by microquasars whose spectral energy distribution peaks in the MeV-GeV band. This was first proposed by Kaufman Bernad\'o et al. (2002), an then discussed in more detail by Romero et al. (2004) and Bosch-Ramon et al. (2004).     

From a statistical point of view, the variable low-latitude EGRET sources have some characteristics that allow to consider them as a distinctive population. For instance, $\log N - \log S$ studies indicate that these sources have a distribution that is different from both galactic radio pulsars and massive molecular clouds and star forming regions (Bosch-Ramon et al. 2004). This is consistent with the fact that high-mass microquasars have large proper motions (e.g. Rib\'o et al. 2002) and hence they can be found outside their birthplaces, but, due to the limited lifespan of the companion star, they cannot spread out along the galactic plane as much as radio pulsars do.     

Although other types of galactic objects can generate variable gamma-ray emission (see, e.g., Romero 2001), microquasars are probably the most attractive candidates. Let us briefly review how these objects can radiate at very high energies.

\section{Gamma-ray production in microquasars: leptonic models}
\label{sect:lepton}

   \begin{figure}
  \centering
   \includegraphics[width=10cm,angle=-90]{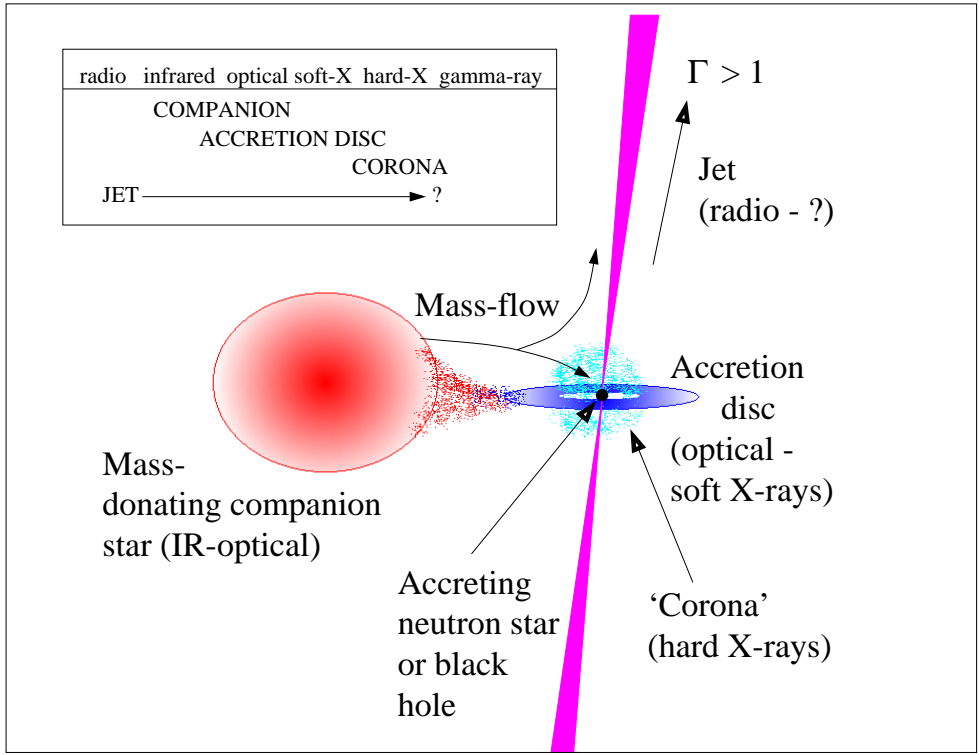}
   \caption{Different components of a microquasars. From Fender \& Macarone (2004).}
   \label{Fig:f3}
   \end{figure}

In Figure 3 we can see all the elements that contribute to the broadband spectrum of a microquasar. We have, in first place, the compact object and the companion star. The star emits from IR to optical (or even up to UV or soft X-rays in the case of very hot stars). In the case of an early-type companion the peak of its thermal emission is around a few eV (UV). Then we have the accretion disk that surrounds the compact object and radiates mainly soft X-rays (its spectrum normally peaks at a few keV). A corona of hot electrons hovers above the inner accretion disk. Its geometry, in the simplest approximation, is spherical. This corona cools by Comptonization of disk photons and contributes with a power-law hard component to the X-ray spectrum. Its high-energy cutoff is observed at a few hundred keV. All these contributions to the total spectrum, shown in Figure 4 for a system like Cygnus X-1, are of thermal nature. In addition, we have the non-thermal contributions from the jet. These contributions will depend on the matter content of the jet, on the energy of its particles, on the strength of the magnetic field, on the geometry of the jet, and on external factors.

Since we observe synchrotron emission from the jet at radio wavelengths, we can assume for sure that there are relativistic electrons and/or electron-positron pairs in the flow. If these electrons are very energetic (say with Lorentz factors up to $\sim10^6$), the synchrotron emission can extend up to hard X-rays with equipartition fields (Markoff et al. 2001). If adiabatic losses are taken into account and lower magnetic fields are adopted ($\sim 200$ G), the synchrotron cutoff drops to $\sim 10$ keV (Bosch-Ramon et al. 2004). However, as pointed out by Atoyan \& Aharonian (1999), SSC losses can be very important and contribute to the high-energy spectrum. Bosch-Ramon et al. (2004), for instance, get SSC luminosities of $\sim 10^{34}$ erg s$^{-1}$ in the EGRET energy range (30 MeV -- 20 GeV), for an inhomogeneous microquasar jet model where SSC losses dominate.

%
%
  \begin{figure}
   \centering
   \includegraphics[angle=0,width=8cm]{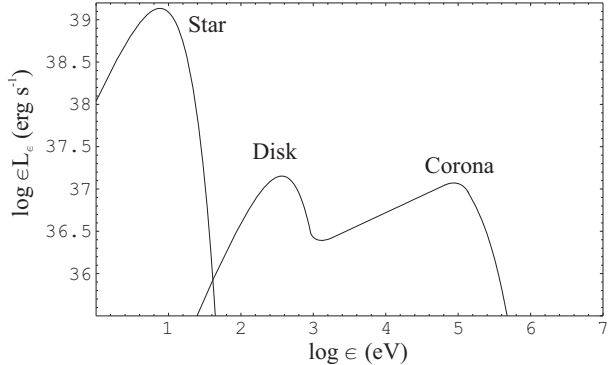}
   \caption{External photon fields to which the jet is exposed in a microquasar with a high-mass companion. }
   \label{Fig:f4}
   \end{figure}

The interaction of the leptonic component in the jet with stellar, disk, and corona photon fields have been considered by Georganopoulos et al. (2002) and Romero et al. (2002) in the context of cylindrical jets. The former authors impose a low-energy cutoff to the electrons in order to reproduce the hard X-ray spectrum of sources similar to Cygnus X-1 as the result of the IC interactions of the jet with the stellar and disk photons. The latter authors adopt the existence of the corona and also consider the interaction of the jet with its field, getting emission at MeV energies, which has been observed in Cygnus X-1. In this work, it is also shown that effects of photon-photon absorption can be important in the inner jet.   

More realistic leptonic models for gamma-ray emission have been recently presented by Bosch-Ramon \& Paredes (2004a, b) and Bosch-Ramon et al. (2004). These models include both SSC and external Compton losses, adiabatic losses, Klein-Nishina calculations, and other improvements. In Figure 5 it is shown a spectral energy distribution obtained for a high-mass microquasar whose inner jet has a magnetic field of $\sim 200$ G at its base (which is located at $\sim 50$ gravitational radii from the compact object). The cutoff for the initial energy distribution is at Lorentz factors of $\sim10^4$ in this case and the fraction of the accreting power that goes to the leptons is $\sim10^{-3}$. 

\begin{figure}
   \centering
   \includegraphics[angle=0,width=10cm]{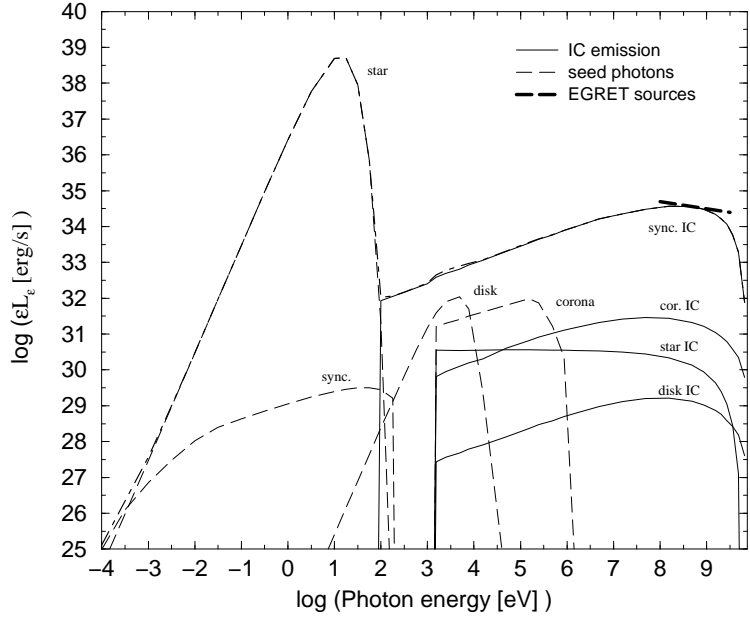}
   \caption{Spectral energy distribution for a microquasar with an inhomogeneous leptonic jet, dominated by SSC losses. External Compton contributions from interactions between the jet and the stellar, disk and corona photon fields are also shown. From Bosch-Ramon et al. (2004).}
   \label{Fig:f5}
   \end{figure}

Jets with very energetic electrons can produce TeV gamma-rays through the upscattering of external photons. In   
Figure 6 we show the spectral energy distribution calculated by Romero et al. (2004) for a microquasar with a high-mass stellar companion (O9 I star). Notice that at TeV energies the spectrum is very soft since the IC interactions occur in the full Klein-Nishina regime. The spectrum also has a break at MeV energies due to the effect of the losses in the electron energy distribution. Some of the unidentified MeV sources detected in the galactic plane by COMPTEL (Zhang et al. 2002, 2004) might correspond to sources of this type, and they are potential targets for high-energy Cherenkov imaging telescopes like HESS or MAGIC. 

\begin{figure}
   \centering
   \includegraphics[angle=0,width=10cm]{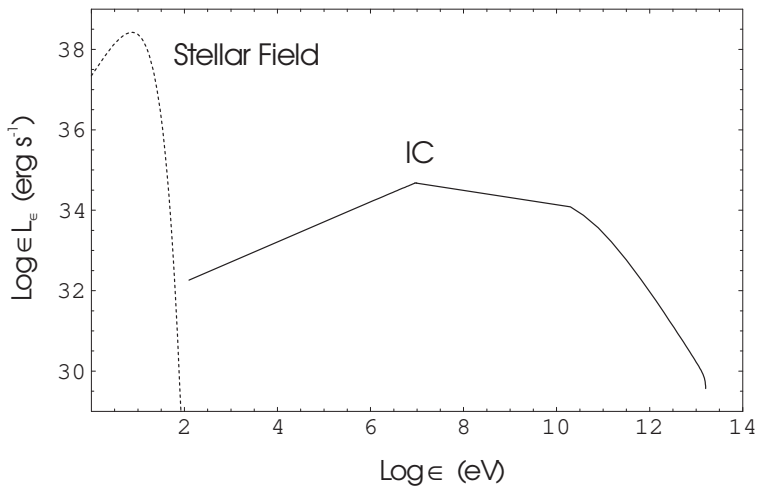}
   \caption{Inverse-Compton spectral energy distribution for a microquasar with a massive (O9 I) stellar companion and a cylindrical inner jet. Leptons in the jet are assumed to have a power-law energy distribution with an index $\alpha=2$ and a high-energy cut-off at multi-TeV energies. Notice the softening of the spectrum at high energies due to the Klein-Nishina effect and the break at MeV energies. From Romero et al. (2004).}
   \label{Fig:f6}
   \end{figure}

Very high-energy emission could also be expected form the extended lobes formed by the deceleration of the jets in the interstellar medium (Aharonian \& Atoyan 1998). In these regions, strong shocks are formed as the result of the interaction of the jet with the ambient matter. These shocks can isotropize and re-accelerate the electrons leading to X-ray synchrotron emission (Corbel et al. 2002) and TeV IC gamma-rays caused through interactions with the 2.7 K photons of the cosmic microwave background. The recent detection of an unidentified TeV source of extended nature in Cygnus region led to the speculation that it might be due to the terminal shock of the jet of Cygnus X-3 (Aharonian et al. 2002). These terminal shocks in microquasars might also be sites of cosmic ray production (Heinz \& Sunyaev 2002).     

In microquasars with low-mass companions the main source of gamma-ray emission should be SSC interactions, because of the paucity of the external photon fields (Romero et al. 2004, Kaufman Bernad\'o 2004, Grenier et al. 2004). The fact that low-mass microquasars are old sources that can present large proper motions (Mirabel et al. 2001, Mirabel \& Rodrigues 2003) makes them attractive candidates to explain the unidentified gamma-ray sources in the galactic halo (Grenier et al. 2004).

\section{Gamma-ray production in microquasars: hadronic models}
\label{sect:hadron}
The matter that accretes onto the compact object in microquasars is formed by both leptons and hadrons. If relativistic protons are present in the jets, whose power seems to be correlated with the accreting power, then we can expect gamma-rays and neutrinos from hadronic interactions. These interactions, however, cannot be among hadrons in the jet, since the particle density is not high enough. An external matter field is necessary. The source of this matter, in the case of systems with high-mass companions, is the associated stellar wind. Early type stars can lose a significant part of their masses through powerful winds, with mass-loss rates up to $10^{-5}$ $M_{\odot}$ yr$^{-1}$. These winds can have terminal velocities of several thousand kilometers per second and very high densities. Romero et al. (2003) have calculated the hadronic gamma-ray emission from high-mass microquasars, finding luminosities similar to those of the EGRET sources at low galactic latitudes for a wide range of reasonable parameters. The results, however, depend significantly on the viewing angle, being favored systems with jets forming small angles with the line of sight. 

\begin{figure}
   \centering
   \includegraphics[angle=0,width=7cm]{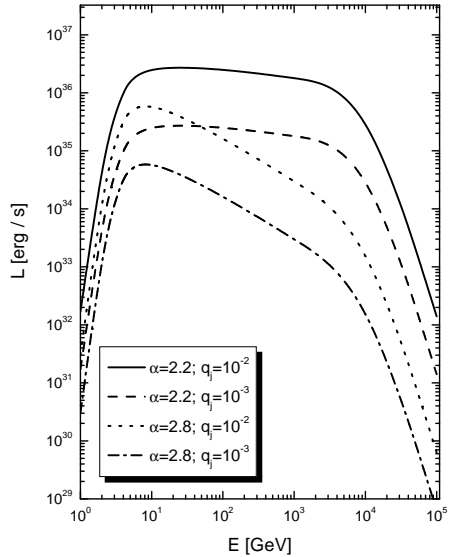}
   \caption{Hadronic gamma-ray spectral energy distribution for a microquasar with a massive (O9 I) stellar companion. Different curves correspond to different jet/disk coupling constant $q_{\rm j}$ and proton spectral index $\alpha$. The viewing angle of the jet is 10 degrees. From Romero et al. (2003).}
   \label{Fig:f7}
   \end{figure}

Romero et al.'s (2003) model assumes that the jet is perpendicular to the orbital plane. In this way the wind impacts onto the jet from the side, allowing the diffusion of the target particles in the jet. However, as discussed by Maccarone (2002), the jet could be strongly inclined. In such configurations, the impact of the jet onto the star would not be unlikely (Butt et al. 2003). However, recent calculations (Romero \& Orellana 2004) show that in most cases the ram pressure of the wind could completely balance the jet pressure, producing a strong shock between the compact object and the star. At this shock particles from the jet can be isotropized and re-accelerated. These particles, then, diffuse in the wind producing a gamma-ray source through $pp$ interactions and the subsequent $\pi^0$-decays. The emission, typically at the level of $\sim10^{32-33}$ erg s$^{-1}$ at energies above 1 TeV, is basically isotropic.

Neutrinos are produced in all these models, with luminosities such that for a reasonable range of distances and integration times hadronic high-mass microquasars might be detectable sources with instruments like ICECUBE (see Bednarek et al. 2004).

High-energy gamma-rays and neutrinos can also be the result of $p\gamma$ interactions. Both SSC and strong external X-ray fields (e.g. from the hot corona) can cool extremely relativistic protons through pion production (Levinson \& Waxman 2001). However, the protons should be accelerated up to very high energies ($\sim 10^{16}$ eV) in the inner jet in order to produce a significant flux of multi-TeV neutrinos. It is in no way established that microquasars can accelerate particles up to such energies at the base of the jets. The high neutrino fluxes predicted by Distefano et al. (2002) on the basis of this model for different microquasars are in some cases already ruled out by AMANDA II data (Torres et al. 2004). A strong TeV gamma-ray flux should accompany the neutrino emission, which was not observed by Cherenkov Imaging telescopes like HEGRA. Bednarek et al. (2004) present an additional discussion on the relative efficiency of the hadronic models based on photopion production. 

\section{Discussion}
\label{sect:discussion}

As it was seen in the previous sections, there are different models that predict high-energy emission from microquasars. A fundamental question is then why are there so few known microquasars within the location errors boxes of EGRET gamma-ray sources?. Just two out of the $\sim$15 confirmed microquasars are strong candidates to gamma-ray sources. However, this microquasars share some common properties that can shed some light on the problem. First, the jets of these systems seem to be persistent and very short. Second, both sources are rather weak X-ray emitters, with a featureless power-law continuum. An finally, both sources might have neutron stars, as emphasized recently by Rib\'o et al. (2004). 

If the two likely gamma-ray microquasars have persistent jets, they should also have a high duty cycle for gamma-ray emission, then making easier for an instrument like EGRET, which integrates over viewing periods of a few days to weeks, to detect them. Sources which change their states from soft to hard states have transient jets, whose high-energy emission would be harder to detect. The presence of stronger X-ray fields in the majority of the microquasars could also contribute to quench strong gamma-ray emission, at least within EGRET sensitivity, because of the likely absorption of MeV-GeV radiation by pair production (Bednarek 1993, Romero et al. 2002). 

New microquasar candidates with similar characteristics to LS 5039 and LSI +61 303 are not easy to find within the EGRET location error boxes of unidentified sources. Normally, there can be dozens of weak non-thermal radio sources within $\sim 1$ degree from the center of gravity of a given EGRET location contour. Since the jets are weak and short, VLBA or ATCA observations are necessary to detect them. The large number of sources renders unpractical to observe all potential candidates, hence some selection criterion is necessary. A good approach is to select those non-thermal sources which display significant radio variability on timescales from month to years (Paredes et al. 2004), and then to implement high-resolution observations to find the jets. In any case, it seems not unlikely that a fraction of the unidentified sources could be undetected microquasars like LS 5039.

Beaming could be another important factor for the potential microquasars behind the unidentified sources. With the exception of V4641 Sgr, Cir X-1, and perhaps Cyg X-3, the rest of the known microquasars do not seem to present strong Doppler boosting in their non-thermal emission. Microblazars, however, could appear as a strong high-energy sources with very weak thermal counterparts. If the high-energy emission is dominated by external Compton scattering, it should be more boosted than the radio emission, which is synchrotron in nature (Kaufman Bernad\'o et al. 2002). The lower energy counterparts of gamma-ray microblazars would be, in general, difficult to find. 

Rib\'o et al. (2004) suggested recently that neutron star microquasars with high-mass companions might be more effective to generate gamma-ray sources. This could be due in part to the absence of a strong corona (which might absorb an important fraction of the gamma-rays) or because the particles in the inner jets reach higher energies than in black hole systems. However, detailed models of jet production by accreting neutron stars should be developed before advancing in this sense. There is also the possibility that the gamma-rays might be produced {\sl independently} of the jets, for instance through interactions of the pulsar wind (I.F. Mirabel, personal communication).

\section{Perspectives}
\label{sect:conclusion}

The forthcoming next few years present exciting perspectives from the point of view of the study of high-energy emission in microquasars. New and powerful instruments will become available at both very high energies (MAGIC, HESS, VERITAS) and mid energies (AGILE, GLAST). The former instruments might make a reality very soon the detection of the first TeV microquasar. Detailed determinations of the spectrum at very high-energies, along with multifrequency data from other instruments will allow to establish the nature of the radiation (e.g. leptonic vs. hadronic). Neutrino observations with ICECUBE, ANTARES, and other possible detectors will also be crucial on this respect. The second group of instruments, will solve in part the problem of the nature of the unidentified gamma-ray sources detected by EGRET. It is very likely that some of the sources result to be new microquasars. To establish their multiwavelength characteristics and to understand how they produce the jets and the high energy emission will be, perhaps, the greatest challenges of microquasar astrophysics in the next years.

\begin{acknowledgements}
I want to thank Felix Aharonian, Peter Biermann, Valenti Bosch-Ramon, Jorge A. Combi, Isabelle Grenier, Marina Kaufman Bernad\'o, F\'elix Mirabel, Josep Mar\'{\i}a Paredes, Marc Rib\'o and Diego Torres for stimulating discussions on microqusars along these last years. I also thank Paula Benaglia for a careful reading of the manuscript. This review is dedicated to the memory of my father, Ricardo Esteban Romero, who passed away on July 13th, 2004, while I was writing the manuscript. This work has been supported by Fundaci\'on Antorchas, ANPCyT, and CONICET. Additional support came from SECyT (Argentina) through a Houssay Prize in Astronomy.
\end{acknowledgements}

\label{lastpage}

\end{document}